\documentclass[%
preprint,
 amsmath,amssymb,
 aps,
nofootinbib,
]{revtex4-1}

\usepackage{geometry}
\usepackage[utf8]{inputenc}
\usepackage{amssymb}
\usepackage{amsmath}
\usepackage{amsfonts}
\usepackage{amstext}
\usepackage{array}
\usepackage{graphicx}
\usepackage{braket}
\usepackage{overpic}
\usepackage{xcolor}
\usepackage{setspace}
\usepackage{hyperref}
\begin{document}

%\preprint{APS/123-QED}

\title{Continuous Time Random Walk with correlated waiting times.\\
The crucial role of inter-trade times in volatility clustering}% Force line breaks with \\

\author{Jaros{\l}aw Klamut}
\email{Jaroslaw.Klamut@fuw.edu.pl}
\author{Tomasz Gubiec}
\email{Tomasz.Gubiec@fuw.edu.pl}
\affiliation{Institute of Experimental Physics, Faculty of Physics, University of Warsaw}

\date{\today}% It is always \today, today,

\begin{abstract}
In many physical, social or economical phenomena we observe changes of a studied quantity only in discrete, irregularly distributed points in time.  
The stochastic process used by physicists to describe this kind of variables is the Continuous Time Random Walk (CTRW).
Despite the popularity of this type of stochastic processes and strong empirical motivation, models with a long-term memory within the sequence of time intervals between observations are missing. 
Here, we fill this gap by introducing a new family of CTRWs. 
The memory is introduced to the model by the assumption that many consecutive time intervals can be the same.
%This way, we obtain a 'Continuous' Time Random Walk within Continuous Time Random Walk.
Surprisingly, in this process we can observe a slowly decaying nonlinear autocorrelation function without a fat-tailed distribution of time intervals.
Our model applied to high-frequency stock market data can successfully describe the slope of decay of nonlinear autocorrelation function of stock market returns.
The model achieves this result with no dependence between consecutive price changes.
It proves the crucial role of inter-event times in the volatility clustering phenomenon observed in all stock markets.
\begin{description}
\item[PACS numbers] 89.65.Gh, 02.50.Ey, 02.50.Ga, 05.40.Fb
\end{description}
\end{abstract}

\pacs{Valid PACS appear here}% PACS, the Physics and Astronomy
\maketitle

%\tableofcontents

\section{Introduction}

In recent years we could observe a rapid increase of interest in point processes and their applications \cite{embrechts2013modelling}.
This kind of stochastic process considers events occurring irregularly in time and describes times between these events and their dependencies.
Two of the most popular models are Autoregressive Conditional Duration (ACD) \cite{acd} and Hawkes model \cite{HAWKES1971a, HAWKES1971b}.
The canonical versions of both models include short-range dependencies (for ACD see \cite{acd, Dufour2000, ENGLE2001113, ENGLE1997187, Bauwens2000, Grammig2000, Pacurar2008} , for Hawkes see \cite{HAWKES1973, Ogata1978, Brillinger1975, Oakes1975, MR1950431, Bowsher2007, Chavez2005, Hewlett06clusteringof, Large2007}).
However, both of them have been extended to describe long-range memory (for ACD see \cite{Beran2015, Jasiak1999, Karanasos2008, 10.2307/1391120, BERAN2002393, DEO20103715, deo_hurvich_soulier_wang_2009, zbMATH06012448, Jasiak1998b, Hautsch2012, Bhogal2019}, for Hawkes see \cite{hawkes2018,  Bacry2015, daley2008pointprocessesII, Chavez2012, Bacry2012, bacry2014a, PhysRevE.85.056108, Filimonov2015, Hardiman2013, PhysRevE.90.062807, jaisson2015}).

In many cases, not only do we observe some events occurring irregularly in time, but also a certain value that can be measured in these discrete moments.
The high-frequency transaction data from a stock market is an excellent example.
We observe events - transactions occurring in specific moments, but also we can relate quantities of price and volume with each transaction.
Of course, in such cases, the inter-event times can be modeled as a point process.

\begin{figure}[ht]
\centering
\includegraphics[width=0.66\textwidth]{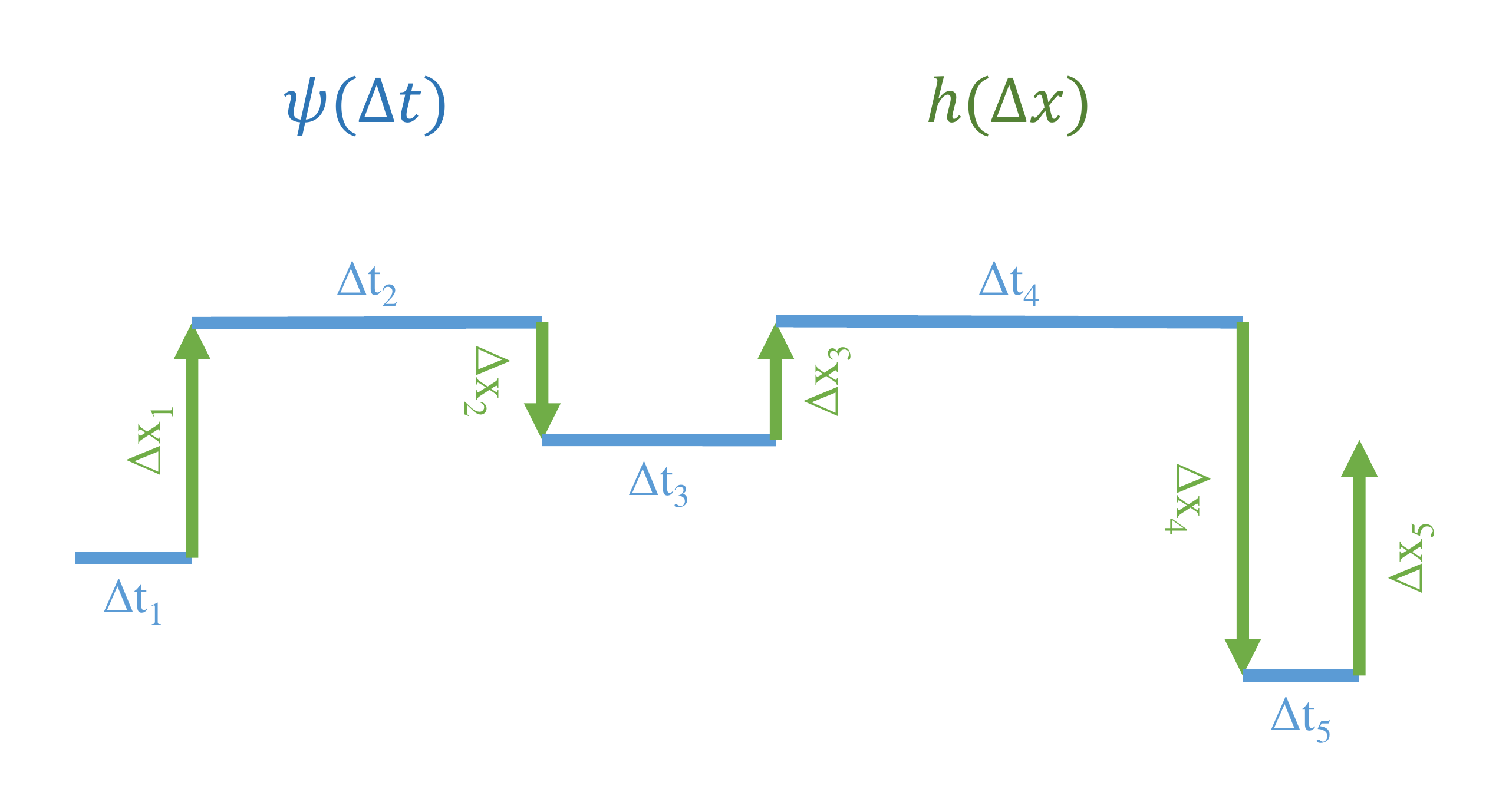}
\caption{The example trajectory of the continuous-time random walk (CTRW) consisting of jumps of process values $\Delta x _n$ preceded by waiting times $\Delta t _n$.
In the canonical CTRW $\Delta t _n$ and $\Delta x _n$ are i.i.d.~random variables drawn from the distributions $\psi(\Delta t _n)$ and $h(\Delta x _n)$ respectively.
In this paper, we consider the CTRW model with long term dependence in the series of waiting times $\Delta t _1, \Delta t _2, \ldots, \Delta t _n$.}
\label{FIG1}
\end{figure}

The first formalism to describe dynamics of observed value changing in distinct, unevenly spaced 'points in time' was continuous-time random walk (CTRW) \footnote{Point processes extended to fit this phenomenon are called marked point processes \cite{MR1950431}.} \cite{montroll1965} proposed in 1965 by Montroll and Weiss.
The CTRW models have found many applications, including astrophysics, geophysics, econophysics, and sociophysics.
For a more detailed review, see \cite{Kutner2017}.
In the canonical CTRW, both increments of the observed process and waiting times (inter-event times) are i.i.d.~random variables.
An examplary trajectory of such a process is shown in Fig.~\ref{FIG1}.
All kinds of random walks, starting with normal diffusion, through anomalous diffusion (both subdiffusion and superdiffusion), to Levy flights, can be described within the  CTRW formalism.
It can be achieved by using specific distributions of waiting times or increments (especially with heavy tails) and by considering memory in waiting times, increments or dependence between them.
The CTRW models with correlated increments were initially proposed to describe lattice gases \cite{kutner1985, kehr1981, HK7879}.
More recently, they have been used to model high-frequency financial data \cite{f1, f2, f3, f4, f5, f6, f7, f8, f9, f10, TG_1, Gubiec2017, Klamut2019}.
On the other hand, the CTRW models with correlated waiting times are not well-studied.
Except for a few papers \cite{metzler2010, PhysRevE.80.031112, f5}, these models were not analyzed nor used to model empirical data.
This fact is surprising, in the light of recent popularity of point processes such as ACD or Hawkes process.
The aim of this work is to fill this gap.
We propose a new CTRW formalism, which is capable of considering dependencies in inter-event times.
The main point is to model long-range memories in a sequence of waiting times.
The formalism proposed in this paper, despite its simplicity, is general enough to explain selected properties of empirical data.

The paper is organized as follows.
In Sec. \ref{SEC_EMP}, we present the exemplary motivation for the model with correlated waiting times based on the financial data.
Next, we propose a way to include dependencies between the waiting times in Sec. \ref{SEC_TIMES}, especially the long-range memory.
Then in Sec. \ref{SEC_CTRW2}, we solve the CTRW model with correlated waiting times, by calculating its propagator, moments and the autocorrelation function (ACF) of increments. 
We also fit our model to tick-by-tick transaction data from the Warsaw Stock Exchange in Sec. \ref{SEC_FIT}.
Finally, we sum up our work in Sec. \ref{SEC_CON}.
Additionally, we include Appendix \ref{SEC_APP} to clarify the mathematical methods used to obtain our results.

\section{Motivation  \label{SEC_EMP}}
%\subsection{Empirical dependencies (in intra-day data) \label{SSEC_NON_STAT}}
%As mentioned in the introduction, the model presented in this paper is directly motivated by high frequency, tick-by-tick data from the stock market \cite{bossa}.
%However, such dependencies as described here can be observed in different areas and their possible applications go far beyond those covered in the paper.
%As mentioned in the introduction,
Models with interdependent waiting times are used for describing electron transfer \cite{Dasenbrook2015}, firing of a single neuron \cite{Karsai2012}, interhuman communication \cite{Wu2010} and modelling of earthquakes \cite{Livina2005, Lennartz2008, Zhang2020, Zhang2020_2}.
An excellent example of a process with correlated inter-event times that we will describe in this manuscript is tick-by-tick transaction price data from the stock market \cite{bossa}.
This data is very convenient to use, as it is high quality and easily accessible in large amounts.

Firstly, let us recall two basic stylized facts observed in the majority of stock markets \cite{RCont2001}. 
%Both of these facts are related to time ACF.
\begin{itemize}
    \item In the ACF of time-dependent log-returns, we observe short-term negative autocorrelation. 
    \item However, for ACF of absolute values of time-dependent log-returns, we observe slowly decaying positive autocorrelation.
\end{itemize}
The latter is considered a reminiscence of the volatility clustering phenomenon.
Usually, the CTRW models used to describe high-frequency stock market data consider waiting times $\Delta t_n$ as inter-transaction times, and process increments $\Delta x_n$ as logarithmic returns between consecutive transactions.
Taking into account the so-called bid-ask bounce phenomenon allowed the CTRW processes to reproduce the first stylized fact of short-term negative autocorrelation \cite{montero2007, TG_0, TG_1}.
In this type of models waiting times $\Delta t_n$ are i.i.d.~variables and only the dependence between $\Delta x_n$ and $\Delta x_{n-1}$ is considered.
%Price increments $\Delta x_n$, which are logarithmic returns between consecutive trades, depend on $\Delta x_{n-1}$.
Unfortunately, models considering only this type of dependencies turned out to be unable to describe time ACF of absolute values of price changes \cite{Klamut2019}.
Technically, it is possible to obtain the CTRW model reproducing both stylized facts, but it requires power-law waiting-time distribution $\psi(\Delta t)$.
However, this solution is not satisfying as we can obtain waiting-time distribution directly from the empirical data of inter-transaction times.
It turns out that this distribution is far from a power-law one \cite{TG_1}.
These results suggest that the source of the second stylized fact is not in the distributions of increments $h(\Delta x)$ and waiting times $\psi(\Delta t)$, but in the dependence between consecutive $\Delta x$ and $\Delta t$.
%As we are interested in construction of a CTRW model accurately describing the second stylized fact, in this paper we introduce dependence, including long range one, in sequence of $\Delta t$ into CTRW model.

\begin{figure}[ht]
\centering
\includegraphics[width=0.49\textwidth]{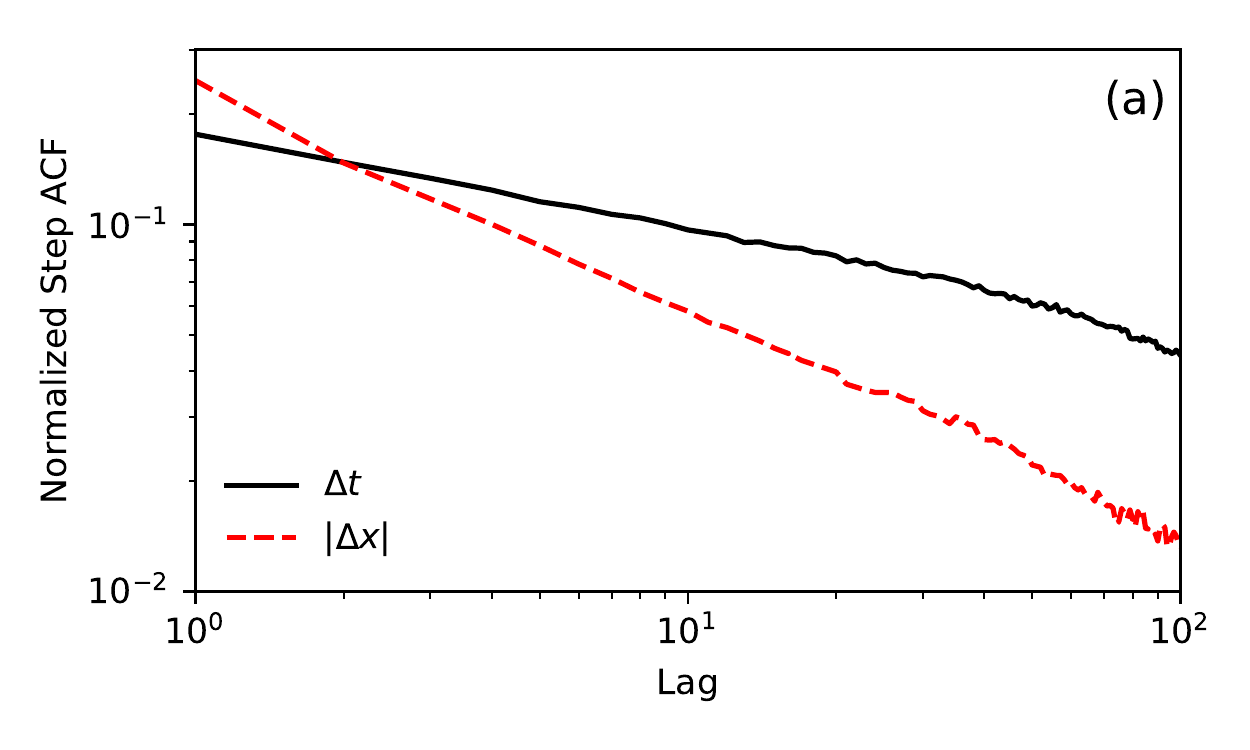}
\includegraphics[width=0.49\textwidth]{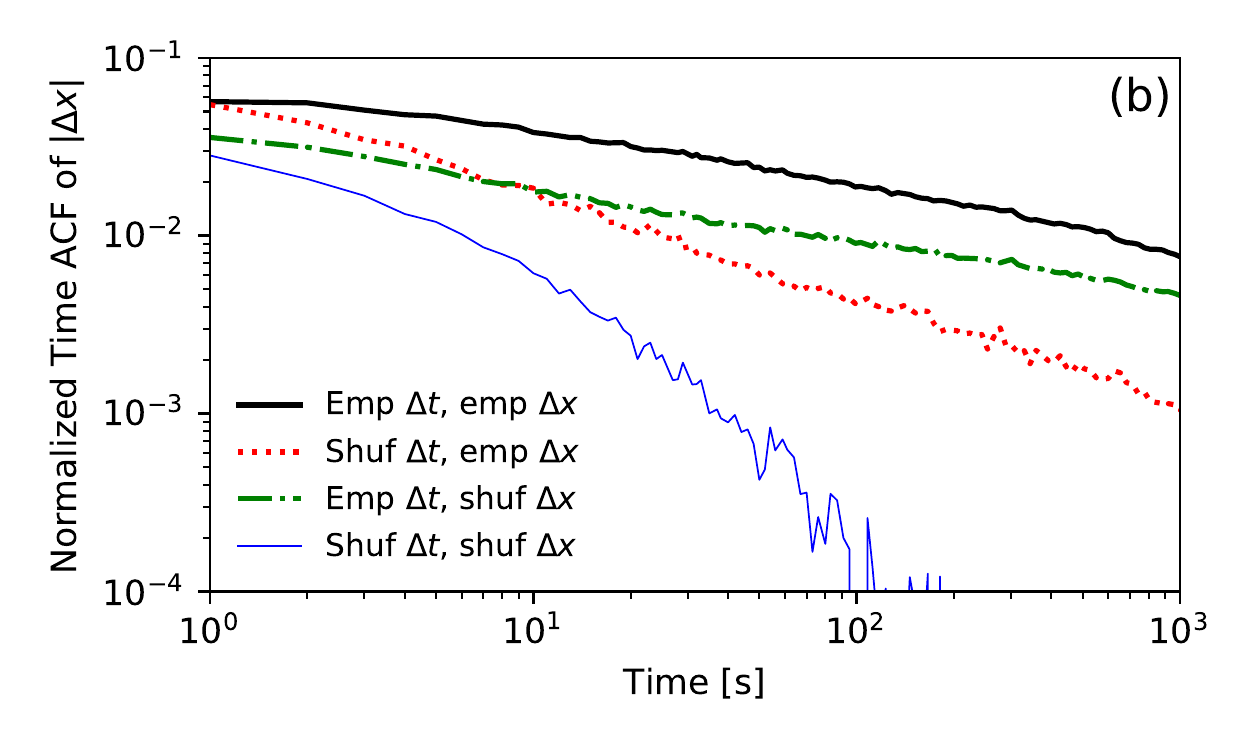}
\caption{Figures \ref{FIG2} and \ref{FIG3} were prepared using transactions data for KGHM (one of the most liquid Polish stock) from period 01.2013 - 07.2017. 
Both figures are on a log-log scale. 
a) The plot of normalized empirical step ACF of $\Delta t$ and $|\Delta x|$. 
Both functions decay like a power-law. 
For lag = 1 autocorrelation of $|\Delta x|$ is higher. 
However, it decays faster, and for long times the memory in waiting times is stronger. 
b) The plot of normalized time ACF of $|\Delta x|$ for 4 time series. 
Presented lines are for empirical data (thick black), empirical price changes and intra-daily shuffled waiting times (dotted red), intra-daily shuffled price changes and empirical waiting times (dash-dotted green), and intra-daily independently shuffled price changes and waiting times (thin blue). 
Considering only empirical dependencies of waiting times reproduces ACF which decays with almost the same slope as the empirical one.}
\label{FIG2}
\end{figure}

Let us start with an empirical analysis of step ACF of series $\Delta t_n$ and $|\Delta x_n|$.
We observe approximately power-law memories in waiting times and absolute values of price changes, see Fig.~\ref{FIG2}a.
For lag $\lesssim 3$ autocorrelation of $|\Delta x_n|$ is higher than autocorrelation of $\Delta t_n$, but for lag $> 3$ it is otherwise. %autocorrelation of $\Delta t_n$ is more significant.
This result suggests that in the limit of long times, the dependence between waiting times may be more critical than dependence between price changes.
To verify this hypothesis we perform a shuffling test.
We compare the time ACF of price changes absolute values for four samples of time series.
The first one is the original time series of tick-by-tick transaction data.
%We expected to observe the second stylized fact in this case.
The second time series keeps the price changes $\Delta x_n$ in the original order but shuffles the order of waiting times $\Delta t_n$.
This way we obtained the time series keeping all dependencies between price changes $\Delta x_n$, but without any dependencies between waiting times $\Delta t_n$.
In the third time series, we kept the original waiting times $\Delta t_n$ but shuffled the price changes $\Delta x_n$.
In the last, fourth time series, both $\Delta t_n$ and $\Delta x_n$ were shuffled.
Let us emphasise that all four time series have the same, unchanged distributions $\psi(\Delta t_n)$ and $h(\Delta x_n)$.
The results are shown in Fig.~\ref{FIG2}b.
As expected, we observe slow, almost power-law decay of time ACF for the first, empirical time series.
Surprisingly, removing dependencies between waiting times does not change the time ACF in the limit of $t \to 0$, but significantly increases its slope of decay in the long-term.
On the other hand, removing dependencies between price changes decreases the time ACF dividing it by an almost constant factor but does not change the slope of the decay.
Removal of all dependencies still leads to positive time ACF, which is a result of the non-exponential empirical distribution of waiting times.

%These observations motivate us to construct the CTRW model with long-range dependencies between waiting times, which should be able to reproduce slowly decaying ACF as in financial data.
%They even suggest that it is impossible to model empirical data without using long-term dependencies in waiting times.
The empirical observations presented above convince us that it is necessary to consider long-range dependencies between waiting times within CTRW to reproduce slowly decaying ACF of price changes absolute values observed in the financial data.

\begin{figure}[ht]
\centering
\includegraphics[width=0.49\textwidth]{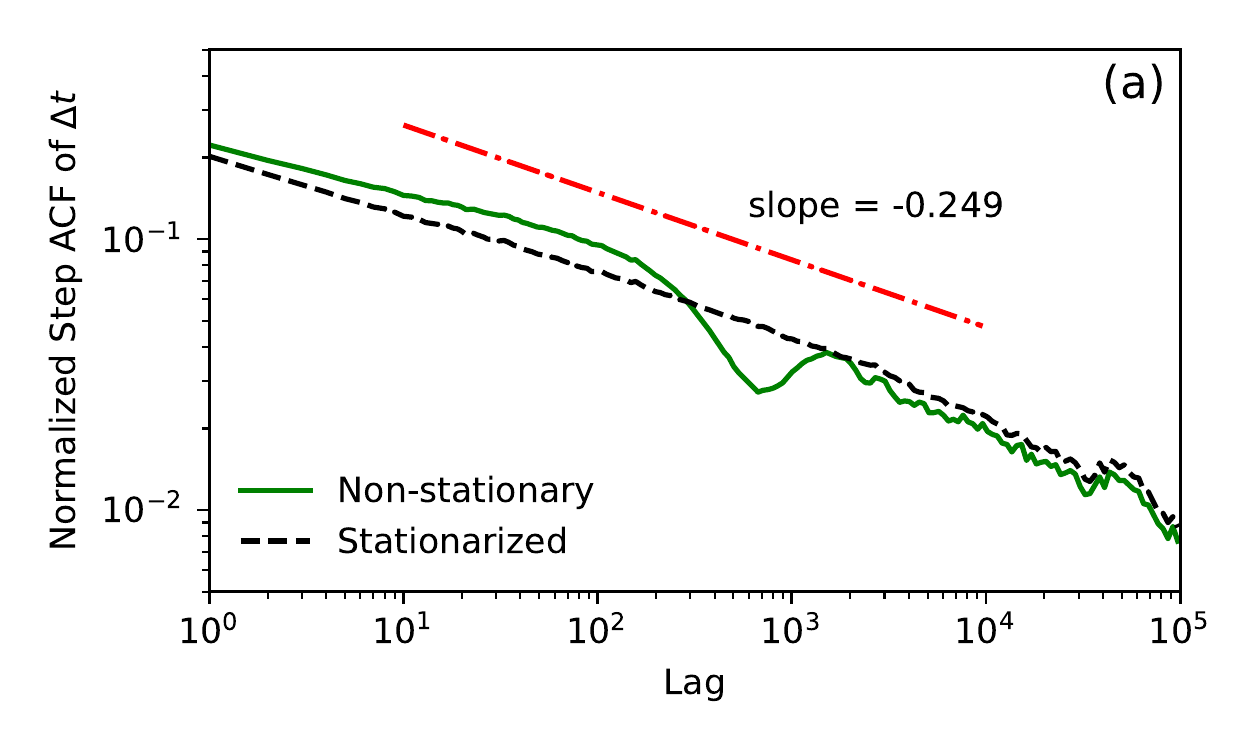}
\includegraphics[width=0.49\textwidth]{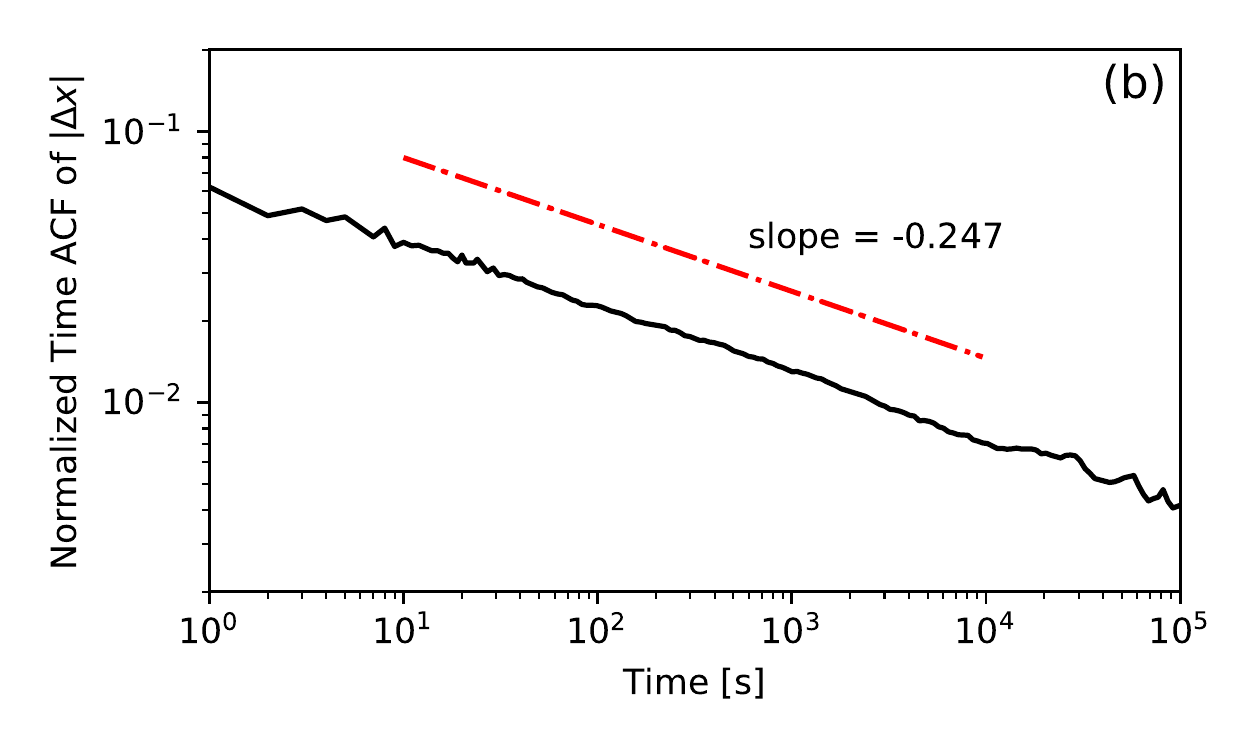}
\caption{All intraday data (waiting times and corresponding price changes) are joined into one data set. 
a) The plot shows normalized step ACF of $\Delta t$ for non-stationary and stationarized case. 
Stationarizing procedure is described in the main text. 
b) The plot of normalized time ACF of $|\Delta x|$ with stationarized waiting times.
Both stationarized autocorrealations decay like a power-law with similar slope.}
\label{FIG3}
\end{figure}

%subsection{Stationary series \label{SSEC_STAT}}
%The results shown above implicate that we should focus on long-term dependencies within waiting times to properly model stock price behavior.
Please note, that in the Fig. \ref{FIG2}, we analyzed step ACF for lags up to 100 and time ACF for times up to 1000 s.
Such limits were chosen due to the length of trading sessions (around 8 hours or 1000 trades).
Unfortunately, these limits are not long enough to detect power-law dependencies.
The only way to increase these limits is by joining all sessions into one sequence.
In this procedure, we merge the end of one session with the beginning of the following one (we omit overnight price changes).
These two periods of the sessions are different, as we observe intraday activity in financial data \cite{TGMW}.
The session begins with short inter-transaction times and a high standard deviation of price changes.
Usually, up to the middle of the session, average inter-trade times increase, and the standard deviation of price changes decreases.
The situation reverts again close to the end of the session.
This phenomenon is called the \emph{lunch effect} \cite{dacorogna2001}.
We use the canonical method to remove intraday non-stationarity by dividing each waiting time by the corresponding average waiting time, depending on the time elapsed since the beginning of the session for each day of week separately \cite{DACOROGNA1993413, Tsay}.
The comparison of step ACF of waiting times for non-stationarized and stationarized data is presented in Fig.~\ref{FIG3}a.
As a result of this procedure, we obtain the power-law decay over four orders of magnitude of lag. 
In Fig.~\ref{FIG3}b, we present the time ACF of price changes absolute values for stationarized data, which also exhibit power-law decay over four orders of magnitude of time lag.
It is now reasonable to ask what is the relationship between decay exponents of these autocorrelations.
Fortunately, the model studied in this paper gives a strict answer to this question.

\section{Process of waiting times \label{SEC_TIMES}}
Let us now focus on the sequence of inter-transaction times $\Delta t_1, \Delta t_2, \ldots, \Delta t_n, \ldots$.
We are looking now for the point process to describe this series, which will be suitable for use in CTRW.
For this reason, we need analytically solvable models.
Moreover, we would like to use the empirical distribution of inter-event times $\psi(\Delta t_n)$ and observe the power-law step ACF, as shown in Fig.~\ref{FIG3}a.
Even these two simple conditions exclude ACD models and Hawkes processes from our considerations.
We are not interested in ACD models, as the power-law ACF can be obtained only within the fractional extension.
In the Hawkes process, both waiting time distribution and autocorrelation depend on the memory kernel. 
Therefore they cannot be set independently. 
By setting the memory kernel, which reproduces the empirical waiting time distribution $\psi(\Delta t_n)$, we obtain specific step ACF, without any degree of freedom to change it.
This feature of Hawkes process significantly hampers its use in the description of empirical data.

\begin{figure}[ht]
\centering
\includegraphics[width=0.66\textwidth]{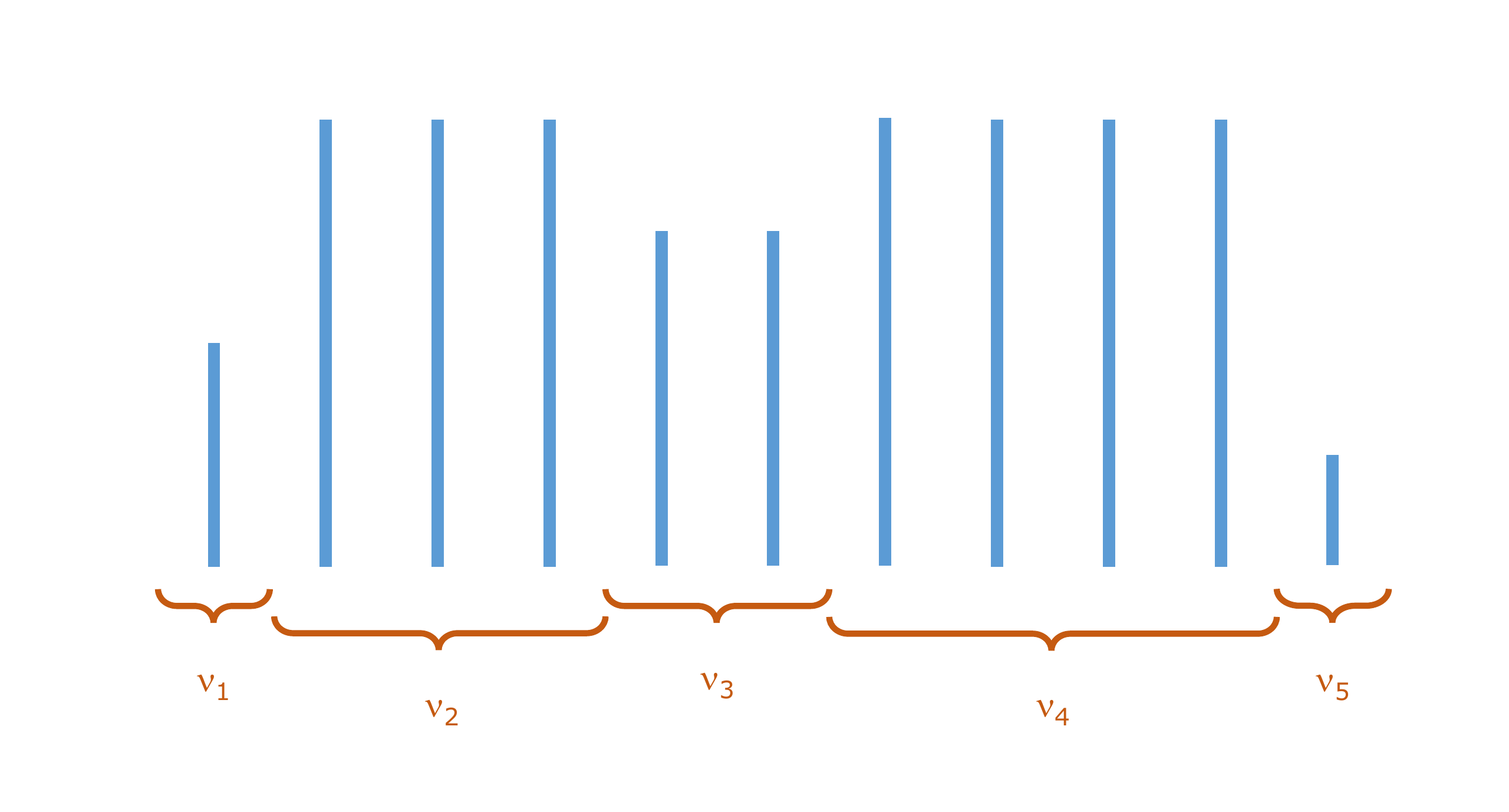}
\caption{The example trajectory of the process of waiting times, whose values correspond to the waiting times $\Delta t_n$ used in the primary CTRW process.
Process values are $\Delta t_1, \Delta t_2, \ldots, \Delta t_n, \ldots$, and they are repeated $\nu_1, \nu_2, \ldots, \nu_n, \ldots$ times respectively. 
Number of repetitions $\nu_i$ are drawn from the distribution $\omega(\nu_i)$.}
\label{FIG4}
\end{figure}

As the solution for our search, we propose a simple point process in which  waiting times $\Delta t_n$ are repeated.
%The solution to our search for a suitable point process turned out to be surprisingly simple.
%We are already dealing with an other stochastic process satisfying our requirements and it is the continuous-time random walk (CTRW).
One can note that the trajectory of such a process is a discrete point process created by evenly sampling the trajectory of a CTRW.
Within the canonical CTRW, values of the process are represented by a spatial variable, and the time is continuous.
Adapting the CTRW to the role of a point process requires the value of the process to represent waiting time and the subordinated time to be discrete.
In the canonical CTRW values of the process are constant during waiting times.
In case of the discrete-time, the analog of waiting time (in canonical case) can be considered as the number of repetitions $\nu_i$ of the same value.
The exemplary trajectory of such adapted, process of waiting times is shown in Fig.~\ref{FIG4}.

We require the waiting times $\Delta t_n$ (values of the process in the discrete subordinated time $n$) to come from distribution $\psi(\Delta t_n)$ ($\Delta t_n > 0$), with a finite mean $\braket{\Delta t}$.
We define $\nu_i$ as the number of repetitions of same waiting times (drawn independently for each series of repetitions).
Let $\nu_i$ be the i.i.d.~random variables with distribution $\omega(\nu_i)$.
In general, it can be any distribution, but to recreate power-law step ACF of waiting times we will focus on fat-tailed distribution with finite first moment $\braket{\nu}$.
In particular, we use zeta distribution with parameter $\rho$
\begin{equation}
\omega_\rho(k) = k^{-\rho} / \zeta(\rho); \quad \zeta(\rho) = \sum_{i=1}^{\infty} i^{-\rho}, \; \rho>1,
\end{equation}
where $\zeta(\rho)$ is Riemann's zeta function.
Its expected value is equal $\braket{\omega} = \frac{\zeta(\rho - 1)}{\zeta(\rho)}$ for $\rho > 2$ and the variance is finite for $\rho > 3$.
The cumulative distribution function is given by $\frac{H_{k,\rho}}{\zeta(\rho)}$, where $H_{k,\rho} = \sum_{i=1}^k i^{-\rho}$ is generalized harmonic number.
Let us introduce $\Omega(k) = \sum_{i=k}^{\infty} \omega(i)$ as a sojourn probability.
We have $\Omega(k) = 1 - \frac{H_{k - 1,\rho}}{\zeta(\rho)}$ for zeta distribution.

%The typical way to describe a stochastic process is to compute its soft propagator $P(x;t | x_0, 0)$.
%It is defined as the conditional probability density that the process value, which was initially (at $t = 0$) in the origin value ($x = x_0$), at time $t$ is equal to $x$.
%The typical way to describe a stochastic process is to compute its soft propagator $P(\Delta t; n | \Delta t_0, 0)$.
%It is defined as the conditional probability density that the process value, which was initially (at $n = 0$) in the origin value ($\Delta t = \Delta t_0$), at time $n$ is equal to $\Delta t$.
%In the case of the process of times, 
We define a soft propagator of the process of times $P(\Delta t; n | \Delta t_0, 0)$, which is the conditional probability density that the waiting time, which was initially (at $n = 0$) in the origin value ($\Delta t = \Delta t_0$), is equal to $\Delta t$ after $n$ steps.
The soft propagator can be expressed by
\begin{equation}
\label{TIMES_P}
P(\Delta t; n| \Delta t_0, 0) = \delta(\Delta t - \Delta t_0) \Omega_1(n) + [1 - \Omega_1(n)] \psi( \Delta t),
\end{equation}
where $\Omega_1(n)$ is sojourn probability obtained from $\omega_1(n)$, which is stationarized distribution of repetition of the first waiting time:
\begin{equation}
\begin{split}
\omega_1(n) &= \frac{\sum_{n'=1} \omega(n + n')}{\sum_{n''=0} \sum_{n'=1} \omega(n'' + n')} = \frac{\sum_{n'=1} \omega(n + n')}{\sum_{n=1} n\omega(n)} = \frac{\sum_{n'=n + 1} \omega(n')}{\braket{\omega}}, \\
\Omega_1(n) &= \frac{\sum_{i = n} \sum_{n'=i + 1} \omega(n')}{\braket{\omega}} = \frac{\sum_{i = 1}  i\omega(i + n)}{\braket{\omega}} = \frac{\braket{\omega} - n \Omega(n + 1) - \sum_{i=1}^{n} i \omega(i)}{\braket{\omega}}.
\end{split}
\end{equation}
The first term of the right hand side of Eq.(\ref{TIMES_P}) is the probability, that the process value will stay constant (equal $\Delta t_0$) after $n$ jumps. 
The second term indicates that there will be a process value jump with probability $1 - \Omega_1(n)$, so new process values will be completely independent, drawn from the distribution $\psi(\Delta t)$. 

Restricting ourselves to $\omega(n)$ in the form of zeta distribution we can obtain 
\begin{equation}
\Omega_1(n) = 1 - \frac{n}{\braket{\omega}} + \frac{n H_{n, \rho}}{\zeta(\rho-1)} - \frac{H_{n, \rho - 1}}{\zeta(\rho-1)},
\end{equation}
and hence the propagator given by Eq.(\ref{TIMES_P}).
The step autocovariance of waiting times $\Delta t_n$ can be expressed as
\begin{equation}
cov(n) = \braket{\Delta t_i \Delta t_{i + n}} - \braket{\Delta t_i}\braket{\Delta t_{i + n}} = \braket{\Delta t_i \Delta t_{i + n}} - \braket{\Delta t}^2,
\end{equation}
where symbol $\braket{\ldots}$ means taking the average.
Note that $\Delta t_{i + n}=\Delta t_i$ with probability $p = \Omega_1(n)$.
With probability $1-p$, the $\Delta t_i$ is independent. 
This leads to
\begin{equation}
cov(n) = p \braket{\Delta t^2} + (1-p) \braket{\Delta t}^2 - \braket{\Delta t}^2 =  \sigma^2_{\Delta t} p =  \sigma^2_{\Delta t} \Omega_1(n).
\end{equation}
We are interested in the asymptotic form of autocorrelation for $n \gg 1$.
We can use following approximation (Theorem 12.21 from \cite{Apostol:1976:IAN})
\begin{equation}
\zeta(\rho) - H_{n,\rho} \approx \frac{n^{1-\rho}}{\rho - 1}.
\end{equation}
Finally, we obtain normalized step ACF
\begin{equation}
corr(n) = \frac{cov(n)}{cov(0)} \approx \frac{n^{-(\rho - 2)}}{\zeta(\rho - 1) (\rho - 2) (\rho - 1)}.
\end{equation}
The step ACF of waiting times decays like a power-law and the decay exponent is $\rho - 2$.
It is worth emphasizing that even considering only $\rho>2$, required for the existence of a finite average number of repetitions, we can obtain any value of the decay exponent.

\section{The primary process \label{SEC_CTRW2}}
Now we are ready to define the primary CTRW process with repeating waiting times.
This process is characterized by the two key properties:
\begin{itemize}
\item changes of the process value $\Delta x_n$ are i.i.d.~random variables from the distribution $h(\Delta x)$, with finite variance $\sigma_x^2$ (and thus finite first two moments $\mu_1$ and $\mu_2$),
\item waiting times $\Delta t_n$ come from the process described in the previous section \ref{SEC_TIMES}.
\end{itemize}
%This way we obtain in some sense a CTRW in CTRW.
%Of course, this name is not a strict one, as the subordinated time is in fact discrete.
Note, that we do not assume any dependence within the series of consecutive changes of the process value $\Delta x_1, \Delta x_2, \ldots, \Delta x_n $.
We do not make any further assumptions about the shape of distributions $h(\Delta x)$.
We managed to obtain the soft propagator of the primary CTRW process and the characteristics derived from it.
%Although the mathematical methods of finding the propagator are exciting, they are not crucial for understanding the main result of this manuscript.
The details of calculations can be found in Appendix \ref{SEC_APP}.
Here we present selected results, namely first two moments and time autocorrelation of changes, in the limit of long times ($t \to \infty$). 
We consider analytical terms ($t, t^2, t^3, \ldots$) and the most significant power-law term when $\rho$ is non-integer.

Using results from the Appendix \ref{SEC_APP}, the first moment of the process for $t \to \infty$ can be approximated as
\begin{equation}
\label{EQ_M1}
m_1(t)  = \mathcal{L}^{-1} \left[-i \frac{\partial \tilde{P}(k;s)}{\partial k} \Big|_{k=0} \right](t) \approx \frac{\mu_1}{\braket{\Delta t}} t + \mu_1 \frac{\alpha \{ \psi \} }{\Gamma(4 - \rho)} t^{3 - \rho}, \quad \rho \in (2;4),
\end{equation}
where $\mathcal{L}^{-1}[\cdot](t)$ is inverse Laplace transform, $\tilde{P}(k;s)$ is the propagator of process in Fourier-Laplace domain, $\Gamma(\cdot)$ is Euler's gamma function and $\alpha \{ \psi \} $ is unknown functional.
The most important term is typical, linear behavior, but we observe additional power-law term.
The second moment can be written in the form
\begin{equation}
\begin{split}
m_2(t) &= \mathcal{L}^{-1} \left[ - \frac{\partial^2 \tilde{P}(k;s)}{\partial k^2} \Big|_{k=0} \right](t) \\
&\approx \mu_1^2 \left( \frac{t}{\braket{\Delta t}} \right)^2  + \sigma_x^2 \frac{t}{\braket{\Delta t}} + \mu_1^2 \beta \{ \psi \} \frac{t}{\braket{\Delta t}} + \mu_1^2 \frac{\gamma \{ \psi \} }{\Gamma(5 - \rho)} t^{4- \rho}, \quad \rho \in (2;5),
\end{split}
\label{EQ_M2}
\end{equation}
where $\beta \{ \psi \}, \; \gamma \{ \psi \}$ are unknown functionals of $\psi$.
From the first two moments of the process, we calculate the process variance (still considering only analytical and the most important power-law term)
\begin{equation}
\label{EQ_S}
\sigma^2(t) = m_2(t) - m_1^2(t) \approx \left( \sigma_x^2 + \mu_1^2 \beta \{ \psi \} \right) \frac{t}{\braket{\Delta t}} + \mu_1^2 \frac{\gamma \{ \psi \} }{\Gamma(5 - \rho)} t^{4- \rho}, \quad \rho \in (2;5).
\end{equation}
It is worth to mention, that for variance the power-law term from the second moment is more important than power-law term from the first moment. 
We can observe normal diffusion for $\rho > 3$. However, there is superdiffusion in case of $\rho \in (2;3)$. 
We obtain ballistic diffusion in the limit $\rho \to 2$.

Having the first two moments, one can calculate ACF of changes for stationary process
\begin{equation}
\label{EQ_C}
C(t) = \frac{1}{2} \frac{\partial^2 m_2(t)}{\partial t^2} - \left( \frac{\partial m_1(t)}{\partial t} \right)^2 \Rightarrow C(t) \approx \mu_1^2 \frac{1}{\Gamma(3 - \rho)} \kappa \{ \psi \} t^{2- \rho},
\end{equation}
where $\kappa\{ \psi \} =  \left( \frac{\gamma \{ \psi \}}{2} - \frac{2 \alpha \{ \psi \}}{\braket{\Delta t}}\right)$, for $\rho \in (2;4)$.
In the limit of $t \to \infty$ and $\mu_1 \neq 0$ we observe a power-law decay of ACF with the exponent $\rho -2$. 
In the case of $\mu_1 = 0$ it can be proved that this exponent is $\rho -1$, so the decay is faster \eqref{AP_MOMENTS}.

It is crucial to emphasize that in Eqs. \eqref{EQ_M1}~--~\eqref{EQ_C} for $\rho$ exceeding the mentioned range, there is still a power-law term with the same dependence on $\mu_1$ and the same time exponent. 
However, the dependence of the amplitude on $\rho$ takes a different more complex form.
%For $\rho > 4$, there is still power-law dependent on $\mu_1^2 t^{2 - \rho}$, but its amplitude is unknown. 
%GARCH: \cite{BAILLIE19963, Koulikov2003}

\section{Empirical results \label{SEC_FIT}}

Now we use the constructed process to investigate the role of correlated inter-trade times in the volatility clustering effect.
We consider this process as a toy model describing high-frequency financial data.
The value of the process represents the logarithm of the stock price.
We can treat transactions as events that change the price.
Therefore the inter-transaction times correspond to waiting times in our model.
The jumps represent the difference of logarithmic prices of consecutive transactions, which are logarithmic returns \cite{f5}.

The CTRW formalism allows us to obtain the autocorrelation of price returns.
Moreover, the same formalism can be used to obtain the non-linear ACF of absolute increments.
This can be achieved by using different jump distributions $h(\Delta x)$. 
To model the process of price changes in time, we should use symmetric distribution $h(\Delta x)$, as the empirical distribution of returns is.
As a result, we obtain vanishing mean $\mu_1 = 0$ and quickly decaying ACF of returns.
To derive the non-linear ACF of absolute returns, we define the new CTRW process, and by calculating its linear ACF, we obtain the non-linear ACF of price increments.
Following \cite{Klamut2019}, if as $h(\Delta x)$ we use only the positive half of the previous distribution multiplied by 2, we deal with the case of non-zero drift and obtain an artificial, monotonically increasing process.
As $\mu_1 \neq 0$, we obtain slow power-law decay of the autocorrelation of absolute returns, as in empirical results presented as the solid black line in Fig. \ref{FIG2}b.

Since we assumed only one type of memory in our model, introduced by the distribution $\omega(\nu)$, we cannot expect that the model will be able to reproduce exact values of the empirical nonlinear ACF of the absolute returns.
The model, however, should be able to reproduce its slope (as in Fig. \ref{FIG2}b the green dash-dotted line reproduces the slope of the solid black line).
The theoretical slope is obtained analytically and is equal $2- \rho$.
It is worth emphasizing that the slope does not depend on the distribution of price changes $h(\Delta x)$ or waiting times $\psi(\Delta t)$ and is fully determined by the single parameter $\rho$ characterizing distribution $\omega(\nu)$.
This fact significantly simplifies comparison with the empirical data, as we are required to estimate only one parameter $\rho$.
On the other hand, the assumption of repeated waiting time is a technical method introducing memory.
We cannot expect to observe such a phenomenon in the empirical time series.
The parameter $\rho$ is a measure of the memory present in the sequence of consecutive waiting times.
Therefore, we estimate this parameter using the slope of the step ACF of waiting times, which in the model is equal $2- \rho$.
It is surprising an potentially essential fact that the exponent of the decay of the nonlinear time ACF is the same as in the step ACF of waiting times.
This result motivates us to compare these two values for empirical financial data.
Of course, in empirical data we also observe a long-term positive step ACF of $|\Delta x|$, which was not included in our model.
Therefore, we can expect that the slope of time ACF of $|\Delta x|$ should be slightly higher than the slope of step ACF of $\Delta t$.
Since a long-term nonlinear autocorrelation is usually interpreted as a reminiscence of the volatility clustering phenomenon, it is interesting to check what part of the observed volatility clustering effect can be explained only by memory between inter-trade times.
We present results for five most traded stocks from the Warsaw Stock Exchange in Tab.\ref{TAB1} (ordered by the number of transactions) with the average inter-trade time not greater than 30 seconds.
%Dates 03.01.2013 - 14.07.2017 (1621 trading days)
%KGHM - 3096625 trades = 1910 per day = <dt> 14.8 s.
%PEKAO - 1501758 trades = 926 per day = <dt> 30.4 s
%For PEKAO mean dt is 30.4 s and for PGNIG 35.8 s.
\begin{table}[ht]
\centering
\setlength{\tabcolsep}{8pt}
\begin{small}
\begin{tabular}{ |c|c|c| } 
\hline
Company & Step ACF $\Delta t$ slope & Time ACF $|\Delta x|$ slope \\  \hline
KGHM    & $-0.25 \pm 0.04$    & $-0.25 \pm 0.02$ \\  \hline
PKOBP   & $-0.33 \pm 0.08$    & $-0.30 \pm 0.02$ \\  \hline
PZU     & $-0.26 \pm 0.03$    & $-0.28 \pm 0.04$ \\  \hline
PGE     & $-0.33 \pm 0.07$    & $-0.36 \pm 0.03$ \\  \hline
PEKAO   & $-0.33 \pm 0.04$    & $-0.37 \pm 0.04$ \\  \hline
%PGNIG   & $-0.249 \pm 0.034$    & $-0.339 \pm 0.068$ \\  \hline
\end{tabular}
\end{small}
\label{TAB1}
\caption{Table with fitted slopes of empirical stationarized step ACF of waiting times and time ACF of price changes absolute values for five most liquid stocks from WSE. 
The time ACF slopes are close to corresponding step ACF slopes.
The analysis was performed on the tick-by-tick market data from the public domain database \cite{bossa}.
The data covers the period 2013-01-03 till 2017-07-14.
For instance, the data set for KGHM contains 3 096 625 transactions.
}
\end{table}

We see that our model can estimate the slope of time ACF with accuracy around 10 \%.

\section{Conclusions \label{SEC_CON}}

We introduced a new Continuous Time Random Walk (CTRW) model with the long-term memory within a sequence of waiting times.
We use a simple model of repeating waiting times instead of commonly used point processes like the ACD and Hawkes process.
Despite its simplicity, our model of repeating waiting times has a few useful properties.
It is stationary, can be treated analytically, and the distribution of waiting times and memory in its series can be set independently.
%We assume that series of waiting times is describe as the analog of CTRW itself.
%This way, we obtained the 'Continuous' Time Random Walk in Continuous Time Random Walk.
As we observe many phenomena with dependencies between waiting times, possible applications of this family of CTRW models go beyond the exemplary application in financial time series modeling presented in this manuscript.

We applied the proposed model to describe the slope of long-term decay of ACF observed in financial time series.
Although we consider only memory in a sequence of waiting times, we managed to show that long term dependencies in waiting times are crucial in explaining the volatility clustering effect. 
%In the future, the next step should be to include the dependence in a sequence of price changes into the model.
%Clearly, it would improve the results, and probably allow to explain volatility clustering phenomenon better. 
%, assumed for instance, in the GARCH type models \cite{BAILLIE19963, Koulikov2003},
Our result advocates that the dependence between consecutive price changes is not the primary carrier of long-range memory in the volatility clustering phenomenon.

\appendix

\section{ \label{SEC_APP}}
In Appendix, we sketch the solution for calculating the moments of the process in the limit of long times. 
All increments of the process $\Delta x$ are independent, so firstly we will focus only on the number of jumps.
We calculate the probability $P_n(t)$ for $n \ge 0$, which is the probability that it will be exactly $n$ jumps up to time $t$.
$P_0(t)$ can be obtained directly from the definition, as the probability of no jumps in the time $t$ is 
\begin{equation}
P_0(t) = \Psi(t) \Rightarrow \tilde{P}_0(s) = \tilde{\Psi}(s),
\end{equation}
where $\Psi(t)$ is the sojourn probability for the waiting time distribution. 
For $n \ge 1$ the process will be described by the number of series of waiting times $k$, waiting times in each series $t_i$ and the number of repetitions of waiting times in each series $\nu_i$. 
Particularly, equations $\nu_1 + \nu_2 + \cdots + \nu_k = n$ and $\nu_1t_1 + \nu_2t_2 + \cdots + \nu_kt_k \le t$ must hold. 
The soft propagator $P_n(t)$ for $n \ge 1$ can be written as a sum of two parts:
\begin{enumerate}
\item the $k$-th series of waiting times $t_k$ repeated $\nu_k$ times ended before time $t$ and the process is still in the same position (the next waiting time will be from the new series),
\item the process is during the series of WT $t_k$, which was repeated $\nu_k$ times so far, in the time $t$.
\end{enumerate}

For simplicity of notation, let's redefine $\Omega(\nu) = \sum_{n=\nu+1}^{\infty} \omega(n)$.
\begin{equation}
\begin{split}
P_n(t) &= \sum_{k=1}^n \sum_{\substack{\nu_1,\ldots,\nu_k \\ \nu_1 + \ldots + \nu_k = n}} \int \limits_{\substack{t_1,\ldots,t_k \\ 0 < \delta t}} \psi(t_1) \ldots \psi(t_k) \Psi(\delta t)  \omega(\nu_1) \ldots \omega(\nu_k) dt_1 \ldots dt_k \\
&+ \sum_{k=1}^n \sum_{\substack{\nu_1,\ldots,\nu_k \\ \nu_1 + \ldots + \nu_k = n}} \int \limits_{\substack{t_1,\ldots,t_k \\ 0 < \delta t < t_k}} \psi(t_1) \ldots \psi(t_k) \omega(\nu_1) \ldots \omega(\nu_{k-1}) \Omega(\nu_k) dt_1 \ldots dt_k,
\end{split}
\end{equation}
where $\delta t = t - \sum t_i \nu_i$. 
Next, we calculate the Laplace transform ($t \to s$) and Z transform ($n \to z$) to obtain
\begin{equation}
\tilde{P}(z; s) = \tilde{P}_0(s) + \tilde{P}_z(s) = \frac{1}{s} \frac{1}{1 - \tilde{f}(z;s)} \left[ 1 + \tilde{F}(z;s) - z (\tilde{F}(z;s) + \tilde{f}(z;s)) \right],
\end{equation}
where $\tilde{f}(z;s) = \sum_{\nu = 1}^{\infty} z^{-\nu} \tilde{\psi}(s \nu) \omega(\nu)$ and analogically $\tilde{F}(z;s) = \sum_{\nu = 1}^{\infty} z^{-\nu} \tilde{\psi}(s \nu) \Omega(\nu)$. 
Notice that the full soft propagator with included jumps can be easily expressed as the Z transform of $\tilde{P}_n$ at the point ${z = \tilde{h}(k)^{-1}}$
\begin{equation}
\begin{split}
\tilde{P}(k; s) &= \sum_{n=0}^{\infty} \tilde{P}_n \tilde{h}^n(k) = \tilde{P}(z; s) \Big|_{z = \tilde{h}(k)^{-1}} \\
&= \frac{1}{s} \frac{1 + \tilde{F}(\tilde{h}(k)^{-1};s) - \tilde{h}(k)^{-1} (\tilde{F}(\tilde{h}(k)^{-1};s) + \tilde{f}(\tilde{h}(k)^{-1};s))}{1 - \tilde{f}(\tilde{h}(k)^{-1};s)}.
\end{split}
\end{equation}
The first two moments of the process can be calculated as derivatives of the propagator at the point $k=0$:
\begin{equation}
\label{AP_MOMENTS}
\begin{split}
\tilde{m}_1(s) &= -i \frac{\partial \tilde{P}(k;s)}{\partial k} \Big|_{k=0} = \frac{\mu_1}{s} \frac{J_0 + j_0}{1-j_0}, \\
\tilde{m}_2(s) &= - \frac{\partial^2 \tilde{P}(k;s)}{\partial k^2} \Big|_{k=0} \\
&= \frac{2\mu_1^2}{s} \frac{j_1(J_0 + j_0) + (1-j_0)(J_1 + j_1 - J_0 - j_0)}{(1-j_0)^2} \\
&+ \frac{\mu_2}{s} \frac{J_0 + j_0}{1-j_0},
\end{split}
\end{equation}
where we introduced
\begin{equation}
j_n = j(n;s) = \sum_{\nu=1}^{\infty} \nu^n \ \tilde{\psi}(s\nu)\ \omega(\nu), \quad J_n = J(n;s) = \sum_{\nu=1}^{\infty} \nu^n \ \tilde{\psi}(s\nu)\ \Omega(\nu).
\end{equation}
Next, we focus on the specific power-law memory. 
We set the distribution of the number of repeats to be Zipf's distribution with the parameter $\rho$: $\omega(\nu) = \frac{\nu^{-\rho}}{\zeta(\rho)}, \; \rho > 2$. 
The parameter $\rho$ has to be bigger than two because the distribution of the number of the repeats must have finite mean not to break ergodicity. 
Also, we expand the moments into series assuming very small $s$ (so for long times). 
To do that we need expansions of $j(n;s)$ and $J(n;s)$ for $n = \{0,1\} < (\rho - 1)$.
One can express $j(n;s)$ as power-law sum
\begin{equation}
j(n;s) = \frac{1}{\zeta(\rho)} \sum_{\nu = 1}^{\infty} \tilde{\psi}(s\nu) \ \nu^{-(\rho - n)} = \frac{s^{\rho - n - 1}}{\zeta(\rho)} \underbrace{ \sum_{\nu = 1}^{\infty} \tilde{\psi}(s\nu) \ (s\nu)^{-(\rho - n)} \ s}_{I}.
\end{equation}
The behaviour of sum $I$ can be estimated by integrals 
\begin{equation}
\int_{2s}^{\infty} \tilde{\psi}(x)x^{-(\rho-n)}dx < I <\int_{s}^{\infty} \tilde{\psi}(x)x^{-(\rho-n)}dx.
\end{equation}
Therefore, we can approximate sum $I$ into series and finally obtain
\begin{equation}
j(n;s) = C_n s^{\rho - n - 1} + C_n^0 + C_n^1 s + C_n^2 s^2 + C_n^3 s^3 + \cdots.
\end{equation}
One can calculate
\begin{equation}
C_n^0 = j(n;0) = \frac{1}{\zeta(\rho)} \sum_{\nu = 1}^{\infty} \nu^{-(\rho - n)} = \frac{\zeta(\rho - n)}{\zeta(\rho)} \ge 1.
\end{equation}
Moreover we can notice that
\begin{equation}
\frac{C_1^0}{C_0^1} = - \frac{1}{\braket{\Delta t}}.
\end{equation}
Similarly we approximated 
\begin{equation}
J(n;s) = D_n s^{\rho - n - 2} + D_n^0 + D_n^1 s + D_n^2 s^2 + D_n^3 s^3 + \cdots.
\end{equation}
Constant terms are:
\begin{equation}
D_0^0 = \frac{\zeta(\rho - 1)}{\zeta(\rho)} - 1= C_1^0 - C_0^0, \quad D_1^0 = \frac{\zeta(\rho - 2) - \zeta(\rho - 1)}{2\zeta(\rho)}= \frac{C_2^0 - C_1^0}{2}.
\end{equation}
This gives the form of the first moment
\begin{equation}
\tilde{m}_1(s) \approx \frac{\mu_1}{s} \left( C_1^0 + D_0 s^{\rho - 2} \right) \frac{\frac{C_0}{C_0^1}s^{\rho - 2} - 1}{sC_0^1} = - \frac{\mu_1}{s^2} \frac{C_1^0}{C_0^1} - \frac{\mu_1}{s^{4 - \rho}} \frac{D_0 + \frac{C_0C_1^0}{C_0^1}}{C_0^1},
\end{equation}
concerning only terms increasing with time ($s^{-\alpha}, \alpha > 1$): analytical and the most important power-law one.
Switching to time variables, we obtain:
\begin{equation}
m_1(t) = \mathcal{L}^{-1} \left[\tilde{m}_1(s)\right] \approx \frac{\mu_1}{\braket{\Delta t}} t - \mu_1 \frac{D_0 + \frac{C_0}{\braket{\Delta t}}}{C_0^1 \Gamma(4 - \rho)} t^{3 - \rho}.
\end{equation}
The second moment can be expressed as
\begin{equation}
\begin{split}
\tilde{m}_2(s) &\approx \frac{2 \mu_1^2}{\braket{\Delta t}^2}s^{-3} -\mu_1^2 \frac{4 C_0^2 + 3 C_0^1 \braket{\Delta t} + 2 C_1^1 \braket{\Delta t} + 2 D_0^1 \braket{\Delta t} + C_2^0 \braket{\Delta t}^2 }{2 C_0^1 \braket{\Delta t}^2} s^{-2} \\
&- \mu_1^2 \frac{D_0 + C_1 - D_1 + 2 \frac{C_0}{\braket{\Delta t}}}{C_0^1 \braket{\Delta t}} s^{\rho - 5} + \frac{\mu_2}{\braket{\Delta t}}s^{-2}.
\end{split}
\end{equation}
This gives us the variance in the time domain, presented in the main text.

\bibliographystyle{unsrt}
\bibliography{my}{}

\end{document}